# Options of Different Rescue Periods on Transport Tools

Mengjia Zhou, Jiahui Chen, Bernie Liu

**Abstract**

In this paper, we construct a universal model to study the search and rescue work in lost planes. We establish an evaluation and decision model for traffic rescue tools. According to the characteristics of different periods, we utilize the combination of Analytic Hierarchy Process (**AHP**) and Fuzzy Synthetic Evaluation (**FSE**) to assess the capability of rescue tools in different period. Then, combined with the actual situation, determine the selection of rescue tools in different periods.

## 1. Introduction

### 1.1 Problem Background

Since the invention of aircraft, countless people have lost their lives in plane crashes. The probability of survived after crashed is very small. Hence, how to organize effective rescue work after crashing seems more significant. Meanwhile, even if there's no possibility surviving, we should salvage the remains of the victims and find the black box. We are eager to know the truth. Searching is the essence of respect for people. The MH370 event is enough to make people profound reflection the way of search and rescue. In this passage, we utilize the method of mathematical modeling to design a common use plan in searching for the lost plane and recognize different types of plane.

### 1.2 Previous Research

There are many outstanding achievement in existence for searching and rescuing crashed aircraft. Many of their achievements have been put into to use. Providing a good solution for salvaging the aircraft. However, all of these seem to lack of comprehensive consideration. Since MH370 is not found can be seen this problem. Some identify rescue plans did not adequately take into account the differences in the way of participating in search and rescue situations in different period's tools. In determining the rescue program did not take into account the characteristics of the surrounding waters. Some did not even consider how to fly the aircraft will be able to search an area as large as possible in the shortest possible time. But it is undeniable that their research has achieved outstanding results and is very effective in terms of search and rescue aircraft crash.

In this paper, a new approach on how to choose the rescue aids and will be promote the search more efficiently

On the basis of our predecessors, we make-up the models and methods that they not taken into account. We conducted this study to determine whether to choose aircraft, ships and Submarine .Further studies are carried out based on the comparison between model and reality.

### 1.3 Our Work

- According to the characteristics of different periods，we established different plans for rescue tools, search and rescue region. There is no such a program in accordance with the unified model can be closer to the actual situation. Making a different deployment scenarios for different periods of time, making the program more effective search and rescue.
- Establishing evaluation and decision models for different periods of searching and rescue tool choice. You can make the vehicle involved in the rescue of a little more full play to their more efficient search and rescue.

## 2. Symbol Description

●Symbols for evaluation models

| Symbol | Definition |
| --- | --- |
| $V$ | Evaluation grade vector |
| $U$ | Evaluation |
| $R_{ship}$ | Judgment matrix of ship |
| $R_{airship}$ | Judgment matrix of airship |
| $R_{submarine}$ | Judgment matrix of submarine |
| $R_{ij}$ | Line i section j of judgment matrix |
| $CI$ | Consistency Index |
| $CR$ | Consistency proportion |
| $Q_i$ | Weight vector |
| $t$ | arrival time of search and rescue aircraft |
| $m$ | Quality |
| $R$ | Judgment matrix |

●Symbols of fuzzy comprehensive evaluation for sensors

| Symbol | Definition |
|---|---|
| *M* | Factor set |
| *N* | Comments set |
| *W* | Weights |
| *B* | Fuzzy comprehensive evaluation set |

# 3. Options of Different Rescue Periods on Transport Tools

## 3.1 Repeater Service Area

By finding relevant information, we find the following for transport aircraft crash rescue

| 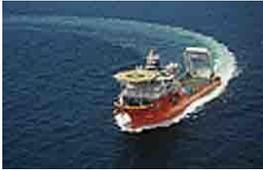 | 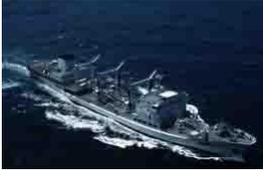 | 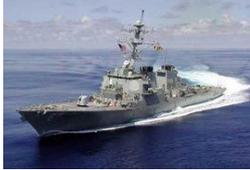 | 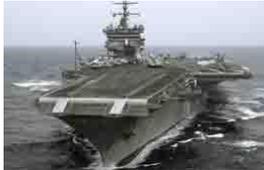 |
|---|---|---|---|
| Professional rescue ship | Large supply ship | Destroyer | Aircraft carrier |
| Rescue equipment professional, complement, rescue personnel experienced | Strong endurance, complete in rescue equipment. | Fast sailing speed | Strong endurance |
| 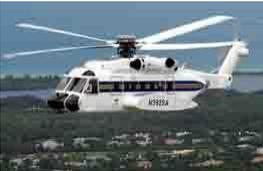 | 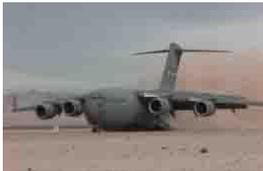 | 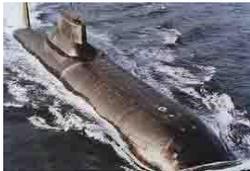 | …… |
| Helicopter | Large transport aircraft | Submarines | Transport name |
| Fast, accuracy | Fast，Large volume | Especially suitable for submarine rescue | Future |

**Figure 3.1** Different types of vehicle

## 3.2 Determination of Evaluation Index

We utilize the different collocation programs on rescue aid in different rescue periods, because of different survival rate and the difficulty of aircraft search. By looking up relevant information , we know that main transports involve in the maritime search and rescue of mainly include the kinds of ships, aircraft and submarines. Ships, which include aircraft carriers, large supply ship, destroyers, small surface ships, professional rescue ships and fishing boats. Aircraft, which include the large transport aircraft, helicopters, fighter, reconnaissance aircraft and antisubmarine aircraft. Submarines, which include nuclear submarine, conventional submarines, small submarines. And evaluation objects are as follows:

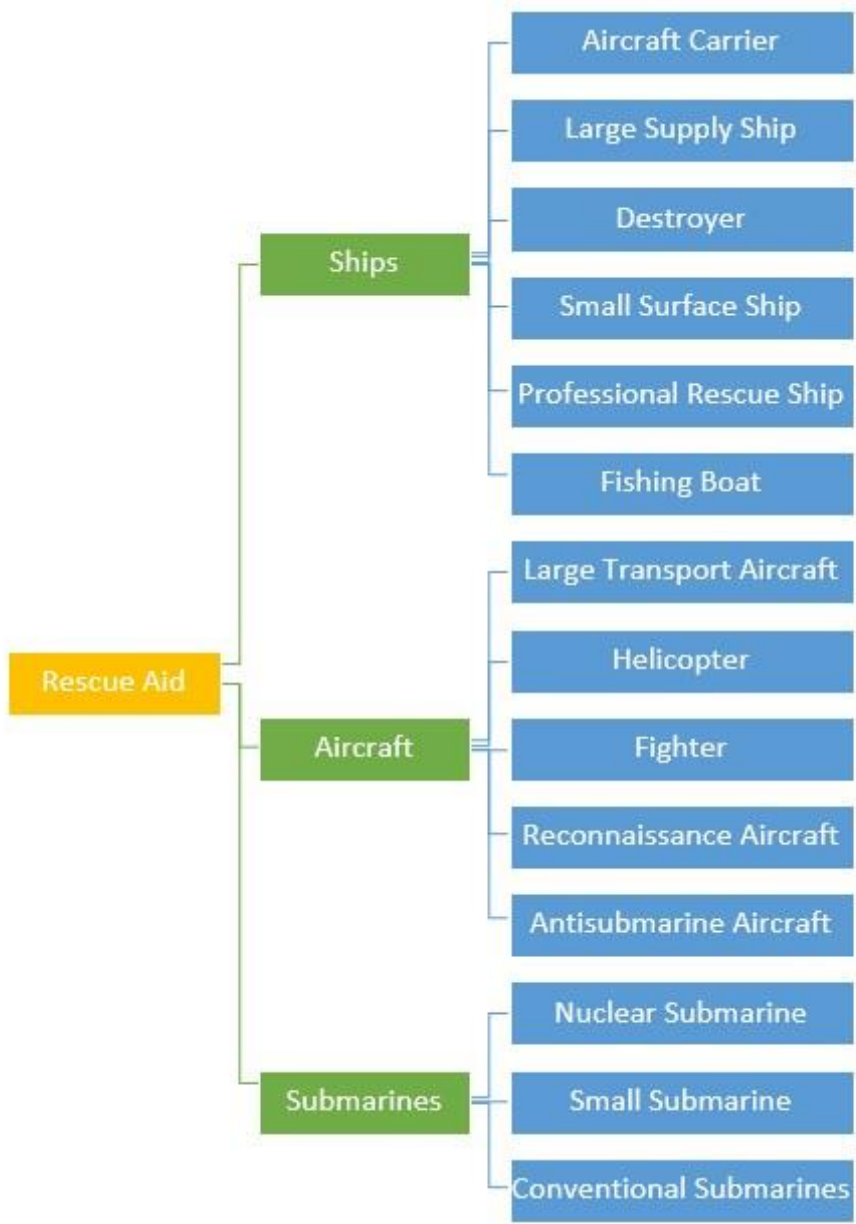

**Figure 3.2** Evaluation Object

## 3.3. Determination of the Weight

First of all, according to the requirements for different rescue periods, we utilize the method of analytic hierarchy process（AHP）to determine the weights of evaluation indexes at different periods. Combining fuzzy comprehensive evaluation method, we can get the score of each transport in different periods and contribution rate, so as to determine the different period of transportation options. The main steps are as follows:

- Step1：Determine the evaluation index

We define that there are i evaluation index: $U= \{u_1, u_2,……,u_i\}$

By comparing the previous search operations in the corresponding period, we reach a conclusion of the indexes for one rescue transport. Those indexes include speed, cost, unit time (s) search area, maximum range, rescue aid, available quantity. The evaluation indexes are presented in Figure 4.3.

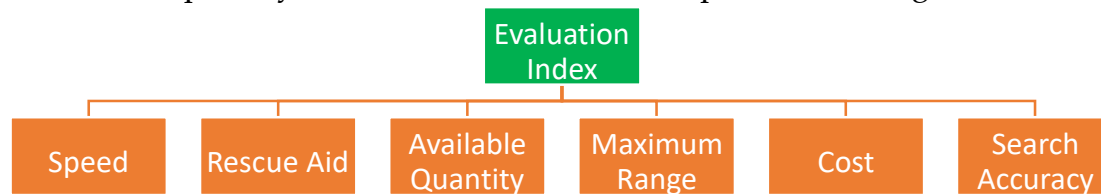

**Figure 3.3** evaluation index

- Step2：Determine the rank of evaluation

We define five levels: Excellent, Good, Moderate, Pass, Fail.

$V=\{v_1, v_2,……v_n\}$

Step3：Establish the fuzzy relationship matrix $R$

After building the rank fuzzy subset, it is necessary to quantify for each factor $u_i(i=1,2,……p)$ one by one. As determining the degree of membership which belongs to evaluated objects on rank fuzzy subset from single factor. We can get the fuzzy relationship matrix R. Through finding out information, the conditions of various rescue tools as indicated in following tables:

**Table 3.1**: The conditions of various rescue tools

(a) The conditions of various ships

| Ship | Speed(*km/h*) |
|---|---|
| Aircraft Carrier | 60.0 40.2 |
| Large Supply Ship | 51.3 28.0 |
| Destroyer | 73.3 |
| Small Surface Ship professional | 20.0 |
| Rescue Ship | |
| Fishing Boat | |

(b) The conditions of various aircrafts

| Aircraft | Speed(*km/h*) |
|---|---|
| Large Transport Aircraft | 500.0 |
| Helicopter | 300.0 |
| Fighter | 2500.0 |
| Reconnaissance Aircraft | 812.0 |
| Antisubmarine aircraft | 907.0 |

| (c) The conditions of various submarines | |
|---|---|
| **Submarines** | **Speed(*km/h*)** |
| Nuclear Submarine | 62.3 |
| Conventional Submarines | 34.8 |
| Small Submarines | 54.9 |

Hence, by the method of quantify analysis, we can get the fuzzy relationship matrix of rescue transports, which as indicated in following:

$$R_{\text{aircraft}} = \begin{bmatrix} 0.8 & 0.2 & 0 & 0 & 0 \\ 0.1 & 0.1 & 0.1 & 0.5 & 0.2 \\ 0.1 & 0.1 & 0.5 & 0.2 & 0.1 \\ 0.1 & 0.3 & 0.5 & 0.1 & 0 \\ 0.7 & 0.2 & 0.1 & 0 & 0 \\ 0.8 & 0.2 & 0 & 0 & 0 \end{bmatrix}; R_{submarine} = \begin{bmatrix} 0 & 0 & 0.2 & 0.6 & 0.2 \\ 0.4 & 0.3 & 0.2 & 0.1 & 0 \\ 0 & 0 & 0.1 & 0.5 & 0.4 \\ 0.8 & 0.1 & 0.1 & 0 & 0 \\ 0.6 & 0.2 & 0.1 & 0.1 & 0 \\ 0.8 & 0.2 & 0 & 0 & 0 \end{bmatrix}; R_{\text{ship}} = \begin{bmatrix} 0.05 & 0.2 & 0.5 & 0.2 & 0.05 \\ 0.5 & 0.3 & 0.1 & 0.1 & 0 \\ 0.1 & 0.6 & 0.1 & 0.1 & 0 \\ 0.5 & 0.3 & 0.1 & 0.05 & 0.05 \\ 0.3 & 0.3 & 0.2 & 0.1 & 0.1 \\ 0.1 & 0.1 & 0.2 & 0.4 & 0.2 \end{bmatrix}$$

In those matrices $R$, element $R_{ij}$(*i-th row, j-th column*) represents the degree of membership from evaluation index $U_i$ mapping to rank fuzzy subset $V_j$. The performance of evaluation index in evaluated objects, which is by means of the fuzzy vector to portray. But other evaluation methods mainly for one practical value of index to portray. Hence, it requires more information to analysis fuzzy comprehensive evaluation from this perspective. Similarly, fuzzy relationship matrix can be drawn in the second layer transports. We do not list them here.

- Step3: By Analytic Hierarchy Process analysis, and according the differences of requirements in golden rescue period, early days of rescue, late days of rescue, we establish judgment matrix and consistency test to determine three kinds of weights. Scale's implication is as follows: **Table 3.2**: Scale's implication

| Scale value | Relationship of two |
|---|---|
| 1 | Both are important |
| 3 | The former is more important than the latter |
| 5 | The former is more important than the latter slightly |
| 7 | The former is more important than the latter strongly |
| 9 | The former is more important than the latter extremely |
| 2,4,6,8 | Represent intermediate state of the adjacent judgment |

### 3.3.1 The Weight of Evaluation Index in Golden Rescue Period

$$\text{Judgment matrix}: S_1 = \begin{bmatrix} 1 & 5 & 3 & 3 & 5 & 3 \\ 1/5 & 1 & 1/3 & 1/2 & 1 & 1/2 \\ 1/3 & 3 & 1 & 1 & 3 & 1 \\ 1/3 & 2 & 1 & 1 & 2 & 1 \\ 1/5 & 1 & 1/3 & 1/2 & 1 & 1/2 \\ 1/3 & 2 & 1 & 1 & 2 & 1 \end{bmatrix}$$

Consistency check: $CI$=0.0083, $CR$=0.0066, the maximum eigenvalue is 6.0414. Hence, the consistency of this matrix can be accepted.

Result: we can get the weight vector is:
$Q_1$=(0.4076,0.0694,0.1665,0.1435,0.0694,0.1435)'. The weight vector of each evaluation index respectively is: Speed=41%, Rescue aid=6%, Available quantity=17%, Maximum range=14%, Cost=7%, Search accuracy=14%.

### 3.3.2 The Weight of Evaluation Index in Early Days

$$\text{Judgment matrix}: S_2 = \begin{bmatrix} 1 & 1/2 & 1/2 & 1/3 & 1/4 & 1/3 \\ 2 & 1 & 1 & 1 & 2 & 1 \\ 2 & 1 & 1 & 2 & 2 & 1 \\ 3 & 1 & 1/2 & 1 & 2 & 1/2 \\ 4 & 1/2 & 1/2 & 1/2 & 1 & 1/3 \\ 3 & 1 & 1 & 2 & 3 & 1 \end{bmatrix}$$

Consistency check: $CI$=0.626, $CR$=0.0497, the maximum eigenvalue is 6.3130. Hence, the consistency of this matrix can be accepted.

Result: we can get the weight vector is:
$Q_2$=(0.0691,0.1885,0.2143,0.1631,0.1206,0.2444)'. The weight vector of each evaluation index respectively is: Speed=7%, Rescue aid=19%, Available quantity=21%, Maximum range=16%, Cost=12%, Search accuracy=24%.

### 3.3.3. The Weight of Evaluation Index in Late Days

$$\text{Judgment matrix}: S_3 = \begin{bmatrix} 1 & 1/5 & 1/2 & 1/4 & 1/4 & 1/5 \\ 5 & 1 & 3 & 1 & 1 & 1 \\ 2 & 1/3 & 1 & 1/2 & 1/2 & 1/3 \\ 4 & 1 & 2 & 1 & 1 & 1 \\ 4 & 1 & 2 & 1 & 1 & 1 \\ 5 & 1 & 1 & 1 & 1 & 1 \end{bmatrix}$$

Consistency check: CI=0.0138, CR=0.0110, the maximum eigenvalue is 6.0690. Hence, the consistency of this matrix can be accepted.

Result: we can get the weight vector is :

$Q_3$=(0.0481,0.2326,0.1096,0.2066,0.2066,0.1965)'. The weight vector of each evaluation index respectively is: Speed=5%, Rescue aid=23%, Available quantity=11%, Maximum range=21%, Cost=21%, Search accuracy=20%. Thus, according to each evaluation index's weight results in different periods, which is as follows chart.

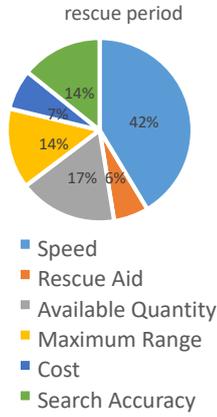

Figure 4.4  Weight of golden rescue period

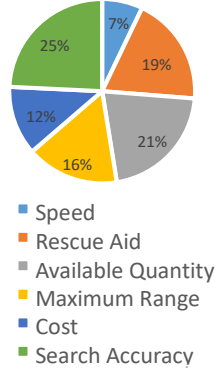

Figure 4.5  Weight of early days

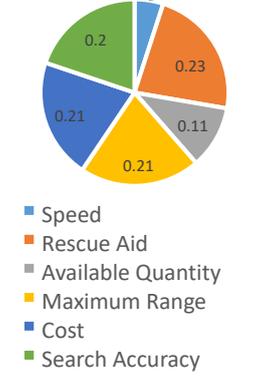

Figure 4.6  Weight of late days of rescue

## 3.4. Different Rescue Scores and Transport Options in Different Rescue Period

The synthesis of the comprehensive evaluation about result vector $B = Q_i \cdot R$

$$Q_i \cdot R = (q_1, q_2, \ldots, q_p) \begin{bmatrix} r_{11} & \cdots & r_{1m} \\ \vdots & \ddots & \vdots \\ r_{p1} & \cdots & r_{pm} \end{bmatrix} = (b_1, b_2, \ldots, b_m) = B \quad (equation\ 5.1)$$

In the above formula, $i$=1,2,3 represents three different rescue period, $q_p$ represents weight of $p_{th}$ index, R is fuzzy relationship matrix. According to the results above, we bring them into the equation can be obtained on the fuzzy comprehensive evaluation result vectors for various rescue transports. And according to the maximum degree of membership principle, can draw various rescue transports to the different level of evaluation indexes. The table below shows different levels which are corresponding to their sores. Adopting

$$F_{last} = F \cdot B \quad (equation\ 5.2)$$

equation can get scores corresponding to the various rescue aids at different times, $F_{level}$ is the different levels of the corresponding points, shown in in the following table:

Table 3.3: Different levels of the corresponding points

| Level | Score($F_{level}$) |
|---|---|
| Excellent | 100 |
| Good | 80 |
| Moderate | 60 |
| Pass | 40 |
| Fail | 20 |

According to the data shown in the above table, calculating the gold rescue period, which can be get different rescue scores in different rescue period in that period.

By the method mentioned above, bringing all the rescue tools in all kinds of data into the model. We can obtain the scores in different indexes:

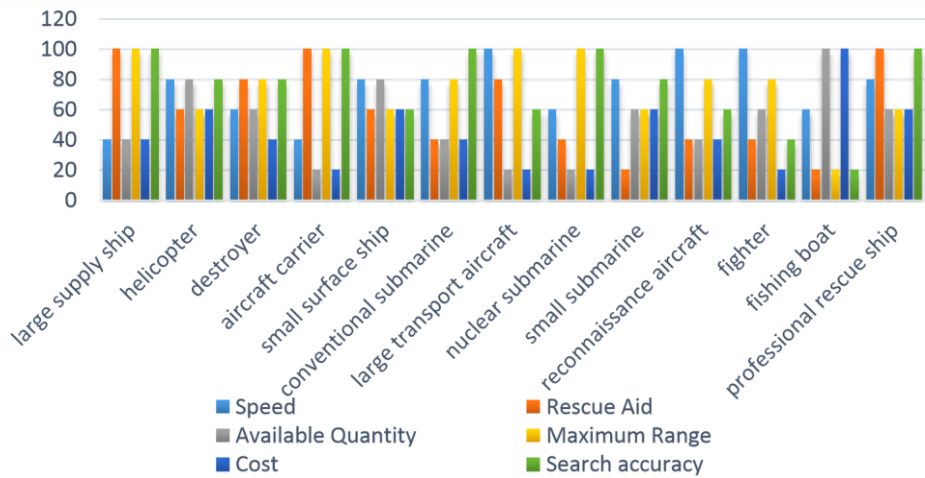

**Figure 3.3** The scores of Vehicle in different indexes

Then we through comparing with actual situation to test model simply. On the basis of figure above, we can see that:
- Aircraft has a large advantage at the evaluation index of speed.
- In terms of search accuracy, the submarine has a large advantage.
- Professional rescue ships have greater advantages at maximum range and rescue equipment.

These all make sense. Hence, it can be a preliminary judge: models and methods is meaningful.

As for the selection and use of final rescue aids in different periods, it can determined by comparing a total score of different transport in gold rescue period.

### 3.4.1. The Scores of Different Tools and Transport Options in

## Golden Rescue Period

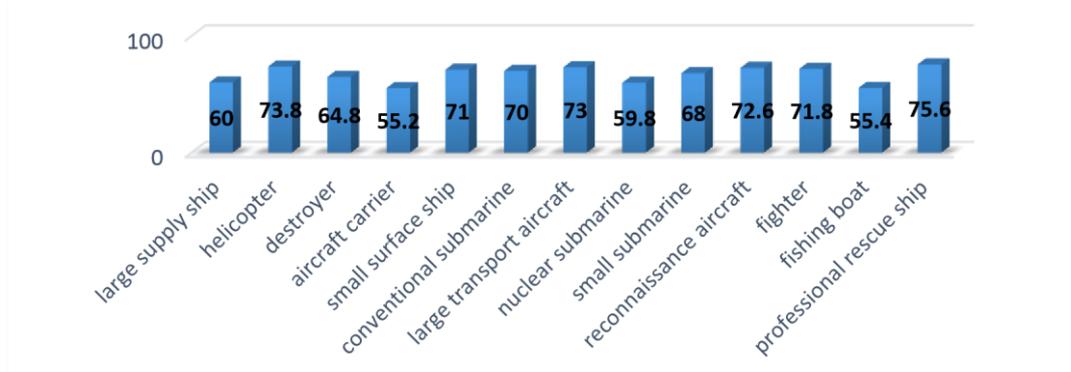

**Figure 3.8** Different traffic tools and their score in gold rescue

In the golden rescue period, due to the chances of survival are hundreds of thousands times of other period, time is the most important evaluation factors. Therefore, it should at all costs to save people's lives. According to the evaluation results that can be seen on the chart, aircraft gets highest scores in all transport rescue tools. Therefore, in the early crash occurred, it should send a large number of aircrafts to the rescue, as much as possible to save more lives. In golden rescue period, we should utilize all rescue aids to come to rescue on the crash site.

### 3.4.2 The Scores of Different Tools and Transport Options in Early Days of Rescue:

● When the crash rescue time has been exceeded gold，crash rescue entered the early stage of aircraft rescue. Evaluation method has taken palace. By the method of above, can get the scores of different tools in early days of rescue.

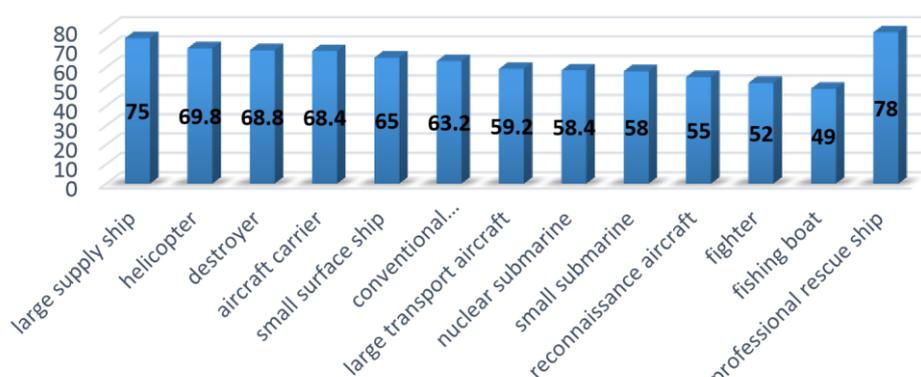

**Figure 4.9** Different traffic tools and their score in the early days rescue

● In the early stage of rescue, the importance of the time factor has decreased because of the golden rescue time has missed. Meanwhile, the degree of importance of each index in the period tends to be close to. The contribution rate of the different modes of transport have changed. According to preliminary search and rescue scores for each transport, taking into account the actual situation. We make the sky, the sea and undersea search integration at the same time. We provide a preliminary search and rescue transport options: large supply ships, helicopters, destroyers, submarines, large transport aircraft, professional rescue ship.

### 3.4.3 The Scores of Different Tools and Transport Options in Late Days of Rescue:

About one month or even longer after the crash, then entered the later stage of search and rescue. At the late stage, the possibility of survival is bare. The wreckages of the plane have moved strongly with the currents occur. Many objects are likely to have sunk to the bottom of the sea. This situation, the degree of importance of maximum range and search accuracy in the period tends to more important. Evaluation standard has taken palace. We can get the scores of different tools in late days of rescue.

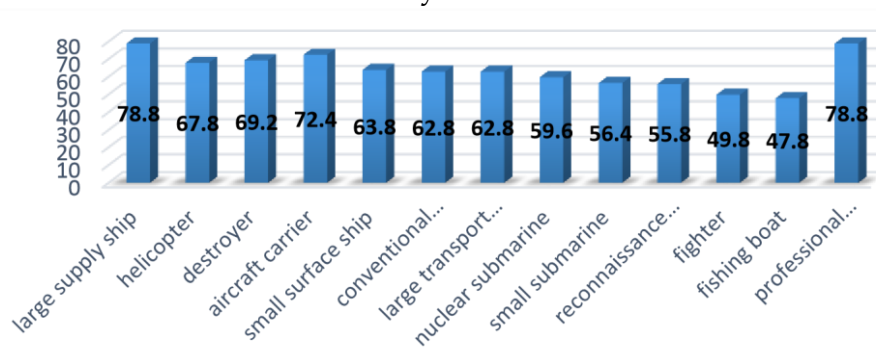

**Figure 4.10** Different traffic tools' score in the early days rescue

According to the evaluation results that can be seen on the chart, we combine principle of sea and air –integrated with Search in the late phase of the actual situation. We provide a late stage search and rescue transport options: large supply ship, helicopter, aircraft carrier, conventional submarine, professional rescue ship.

All in all, we take advantage of combining AHP and fuzzy comprehensive evaluation method which have given above. Combined with the actual data, the different period of rescue aids that are selected, showing in the table below:

**Table 4.4**: Scale's implication

| Search Period | Options of rescue aid |
|---|---|
| Golden Rescue period | All tools |
| Early days of rescue | Large supply ship, helicopter, fighter, conventional submarine, large transport aircraft, professional rescue ship |
| Late days of rescue | Large supply ship, helicopter, aircraft carrier, nuclear submarine, professional rescue ship |

# 4 Model Analysis:

## 4.1 Sensitivity Analysis on the Model

### 4.1.1 Sensitivity Analysis on the Decision Model of Transportation Means

We use fuzzy synthetic evaluation method for the decision model of transportation means. This method is less sensitive when analyzed theoretically. To further prove and analyze the sensitivity of the method in this paper, we changed relevant parameters and data, and obtained the results through comparison. Then we carried out analysis. We changed related parameters or data by using the following methods:

**a).** Amending the quantitative criteria of judgment matrix, mainly changing the parameters to even numbers, and making the uneven numbers at intermediate state
**b).** Amending the judgment matrix, reselecting the judgment matrix
**c).** Adding one or two evaluation indexes or replacing one evaluation index, analyzing the results

When implementing **a)** and **b)**, we find that the changes in the quantitative criteria and numerical value of the judgment matrix are mainly reflected in the changes of weights. Moreover, weights change little (according to the experimental analysis, the changes of weights are all no larger than 15%). However, as we adopt hierarchical scores, unless the numerical value of the judgment matrix changes greatly, otherwise it will hardly affect the final decision result. For **c)**, we add the search index of "search region within unit time". We find that the high-speed transportation means all obtain higher scores. Among the selection schemes for search-and-rescue transportation means during each period, the number of aircrafts increases; for instance, large transport aircraft may appear in the decision scheme in late days of rescue.

We can conclude the following conclusions through the aforesaid experiments: during different rescue periods, the evaluation and decision models of searchand-rescue transportation means in the selection scheme are less sensitive to the selection criteria and slight fluctuation of the numerical values in the judgment matrix. However, they are sensitive to the change of evaluation indexes to some extent. If the evaluation index changes, the result will probably change.

## 4.2 Error Analysis on the Model

### 4.2.1 Error Analysis on the Decision Model of Transportation Means

● The errors of decision model of transportation means mainly reflect in the follows:

① For the subjectivity of weight determination, as the judgment matrix determined by means of analytic hierarchy process is subjective to some extent, its subjective factors are unavoidable even if check consistency is carried out in this paper. Therefore, the weights of the evaluation indexes determined may have some errors caused by subjective factors. (Systemic errors)

② Evaluation indexes are selected incomprehensively or omitted, as a result, the evaluation data may be favorable to a certain transportation means, which may cause errors between the calculation result of the model and the objective real data. (Random error)

③ In fuzzy comprehensive evaluation method, we use different hierarchical scores rather than accurate scores. Moreover, these scores are subjectively determined. Therefore, it may cause errors between the calculation result of the model and the objective real data. (Systemic error)